\documentclass[twocolumn,showpacs,pra,aps,superscriptaddress]{revtex4}

\usepackage{graphicx}
\usepackage{bm}
\usepackage{psfrag}
\usepackage{amsmath}
\usepackage{amssymb}
\usepackage{latexsym}
\usepackage{exscale}
\usepackage{natbib}

\newcommand{\vect}[1]{\mathbf{#1}}
\newcommand{\sprod}{\cdot}

\newcommand{\vprod}{\times}
\newcommand{\trace}{{\rm Tr}}

\newcommand{\dif}{\mathrm{d}}
\newcommand{\ten}[1]{\mbox{\textbf{\textsf{#1}}}}
\renewcommand{\Im}{{\rm Im}\,}
\renewcommand{\Re}{{\rm Re}\,}
\renewcommand{\vec}[1]{\mathbf{#1}}
\newcommand{\mycomment}[1]{}
\newcommand{\Nabla}{\bm{\nabla}}

\newcommand{\D}{\mathrm{d}}
\begin{document}

\title{Impact of amplifying media on the Casimir force}


\author{Agnes Sambale}

\affiliation{Theoretisch--Physikalisches Institut,
Friedrich--Schiller--Universit\"at Jena, Max--Wien--Platz 1, D-07743
Jena, Germany}
\email{agnes.sambale@uni-jena.de}

\author{Stefan Yoshi Buhmann}
\affiliation{Quantum Optics and Laser Science, Blackett Laboratory,
Imperial College London, Prince Consort Road, London SW7 2BW, United
Kingdom}

\author{Ho Trung Dung}
\affiliation{Institute of Physics, Academy of Sciences and Technology,
1 Mac Dinh Chi Street, District 1, Ho Chi Minh city, Vietnam}

\author{Dirk--Gunnar Welsch}
\affiliation{Theoretisch--Physikalisches Institut,
Friedrich--Schiller--Universit\"at Jena, Max--Wien--Platz 1, D-07743
Jena, Germany}

\date{\today}

\begin{abstract}
On the basis of macroscopic quantum electrodynamics, a theory of
Casimir forces in the presence of linearly amplifying bodies is
presented which provides a consistent framework for studying the
effect of, e.g., amplifying left-handed metamaterials on dispersion
forces. It is shown that the force can be given in terms of the
classical Green tensor and that it can be decomposed into a resonant
component associated with emission processes and an off-resonant
Lifshitz-type component. We explicitly demonstrate that our theory 
extends additive approaches beyond the dilute-gas limit. 
\end{abstract}

\maketitle
%
%%%%%%%%%%%%%%%%%%%%%%%%%%%%%%%%%%%%%%%%%%%%%%%%%%%%%%%%%%%%%%%%%%%%%%
%
\section{Introduction}

Among the vast body of literature related to the Casimir forces
\cite{Bordag2001, Milton2004, Lamoreaux2005, Review2006} there is an
almost complete lack of studies concerned with the influence of
amplifying media. This is in stark contrast to the fact that Casimir
forces between amplifying bodies may lead to far-reaching
applications. To name just two examples, amplifying bodies may hold
the key to enhance the impact of novel (meta)material properties on
dispersion forces and to realizing repulsive Casimir forces.

The problem of dispersion forces in the presence of metamaterials has
recently attracted a lot of attention \cite{Kampf2005, Henkel2005,
Tomas2005, Spagnolo2007, Rosa2008a,Sambale2008a,Yang2008}. However,
passive metamaterials suffer from high absorption which restricts
desired metamaterial properties such as lefthandedness
\cite{Veselago1968,Veselago2006} to a narrow spectral bandwidth
\cite{Nistad2008}; this may reduce or completely inhibit an influence
of such properties on dispersion forces. It has been suggested that
the influence of absorption can be mitigated via introducing active
media \cite{Nistad2008}. In practice, gain may be introduced in a
medium by optical parametric pumping \cite{Popov2006} or quantum
cascade lasing techniques \cite{Ginzburg2008, Faist1994}. The
potential of amplifying bodies as a means of realizing repulsive
Casimir forces has been pointed out recently
\cite{Sherkunov2005,Leonhardt2007}. It is already manifest in the
Casimir--Polder (CP) forces acting on individual excited atoms
\cite{Barton70,Wylie1985,Bushev2004,Sambale2008a}. Repulsive
dispersion forces can help to suppress the unwanted phenomenon of
stiction and allows for new classes of nanodevices \cite{Butt2005}.
Note that the repulsive Casimir forces recently measured in
Refs.~\cite{Feiler2008,Munday2009} require the interacting bodies to
be embedded in medium. This is not necessary when realizing repulsive
forces on the basis of amplification. 

A recent calculation of Casimir forces on amplifying bodies
\cite{Leonhardt2007} was based on the assumption that the well-known
Lifshitz-type formula for the Casimir force as an integral over
imaginary frequencies applies without change to amplifying media ---
an approach which neglects excited-state emission processes typical of
amplification. As an alternative, the Casimir force between two dilute
samples of excited atoms has been studied microscopically by summing
over two-atom interactions \cite{Sherkunov2005}. Such an approach is
limited to sufficiently dilute media, whereas a theory nonlinear in
the magnetoelectric properties is indispensable in many applications
such as superlens-scenarios \cite{Podolskiy2005} or anti-stiction
tools \cite{Zhao2003a}.

In this Rapid Communication, we present a macroscopic nonperturbative
theory of Casimir forces on amplifying bodies that presents a
consistent generalisation of additive descriptions based on dispersion
forces on atoms. We establish the consistency of our macroscopic
body--body force with the theoretically \cite{Barton70,Wylie1985} and
experimentally \cite{Bushev2004} well understood force between an
excited atom and a body and show that relevant processes such as
emission by the bodies are fully taken into account. We begin with an
outline of the underlying quantisation scheme \cite{Raabe2008}, which
is an extension of macroscopic quantum electrodynamics (QED) to
amplifying media, and then derive the Casimir force on an amplifying, 
polarisable body of arbitrary shape. Finally, contact to CP forces is
established by applying the general formula to the case of a weakly
polarisable medium.
%
%%%%%%%%%%%%%%%%%%%%%%%%%%%%%%%%%%%%%%%%%%%%%%%%%%%%%%%%%%%%%%%%%%%%%%
%
\section{Quantisation scheme}
Consider an arrangement of polarisable bodies whose linear, local and
isotropic response is described by a spatially varying complex
permittivity $\varepsilon({\bf r},\omega)$ that fulfils the
Kramers--Kronig relations. We allow for bodies that are (linearly)
amplifying in some frequency range [$\Im\varepsilon({\bf r},\omega)
=\varepsilon_I({\bf r},\omega)<0$], assuming the amplifying medium to
be pumped to a quasi-stationary excited state (for details, see
\cite{Raabe2008}). Note that this inversion-type excitation 
is fundamentally different from thermal excitations. In particular, 
the amplifying bodies are explicitly not at (thermal) equilibrium with
their environment.

The quantised electric field can be given as the solution to the
inhomogeneous Helmholtz equation
\begin{equation}
[\bm{\nabla}\times
\bm{\nabla}\times
-\omega^2/c^2\varepsilon({\bf r},\omega)]
\hat{\underline{{\bf E}}}({\bf r},\omega) =
i\mu_0\omega\hat{\underline{{\bf j}}}_N({\bf r},\omega)
\end{equation}
[$\hat{\bf O}({\bf r}) = \int _0^\infty \D \omega
\hat{\underline{\bf{O}}}({\bf r},\omega)+\rm{H.c.}$]
according to
\begin{equation}
\label{edef}
\hat{\underline{{\bf E}}}({\bf r},\omega)
=i\omega \mu_0\int \D ^3 r'\,\ten{G}({\bf r},{\bf r'},\omega)
\cdot \hat{\underline{{\bf j}}}_N({\bf r'},\omega),
\end{equation}
where the classical Green tensor obeys the equation
\begin{multline}
\label{e4}
[\bm{\nabla}\times\bm{\nabla}\times-\omega^2/c^2]
\ten{G}({\bf r},
{\bf r}',
\omega)\\
=\bm{\delta}({\bf r}-{\bf r}')+\omega^2/c^2
[\varepsilon({\bf r},\omega)-1]\ten{G}({\bf r},{\bf r'},\omega)
\end{multline}
together with the boundary condition at infinity. By exchanging
the roles of creation and annihilation operators in the amplifying
space and frequency regime \cite{Scheel1998}, the noise current
density can be given as
\begin{multline}
\label{jn}
\hat{\underline{{\bf j}}}_N({\bf r},\omega)
=\omega\sqrt{\hbar\varepsilon_0\pi^{-1}
 |\varepsilon_I({\bf r},\omega)|}\\
\times\bigl\{\Theta[\varepsilon_I({\bf r},\omega)]
\hat{{\bf f}}({\bf r},\omega)
+\Theta[-\varepsilon_I({\bf r},\omega)]
\hat{{\bf f}}^\dagger({\bf r},\omega)\bigr\}
\end{multline}
[$\Theta$: unit step function, $\Theta(0)\equiv 1$]. The dynamical
variables $\hat{{\bf f}}({\bf r},\omega)$ as introduced in
Eq.~(\ref{jn}) obey bosonic commutation relations such that the
equal-time commutation relation characteristic of the electromagnetic
field holds, $[\hat{{\bf E}}({\bf r}),\hat{{\bf B}}({\bf r'})]
 =i\hbar\varepsilon_0^{-1} \bm{\nabla}\times
 \bm{\delta}({\bf r}-{\bf r'})$, where the electric and induction
fields are given by Eq.~(\ref{edef}) and
\begin{equation}
\label{bdef}
\hat{\underline{{\bf B}}}({\bf r},\omega)
=\mu_0\int \D ^3r'\,\bm{\nabla}\times\ten{G}({\bf r},{\bf r'},\omega)
\cdot \hat{\underline{{\bf j}}}_N({\bf r'},\omega)
\end{equation}
respectively.
The vacuum state $|\bigl\{0\bigr\}\rangle$ of the electromagnetic
field and the partially amplifying electric body is defined by
$\hat{{\bf f}}({\bf r},\omega)|\left\{0\right\}\rangle = \bm{0}$
$\forall {\bf r},\omega$. The quantisation procedure implies that the
Hamiltonian 
$\hat{H}=\hat{H}_++\hat{H}_-
=\int\!\D^3r\!\int _0^\infty\!\D\omega\,
 \hbar\omega\:\{\Theta[\varepsilon_I({\bf r},\omega)]
 -\Theta[-\varepsilon_I({\bf r},\omega)]\}
 \hat{{\bf f}}^\dagger ({\bf r},\omega)\cdot
 \hat{{\bf f}}({\bf r},\omega)
$
generates the correct Maxwell equations via the Heisenberg equations
of motion.
%
%%%%%%%%%%%%%%%%%%%%%%%%%%%%%%%%%%%%%%%%%%%%%%%%%%%%%%%%%%%%%%%%%%%%%%
%
\section{Calculation of the Casimir force}
The Casimir force acting on a partially amplifying body of volume $V$
in the presence of other bodies outside $V$ can be identified as the
average Lorentz force \cite{Raabe2006}
\begin{equation}
\label{e20}
 {\bf F}=\int _{V}\D ^3r\,
\langle \hat{\rho}({\bf r})
\hat{{\bf E}}({\bf r'})
+\hat{{\bf j}}({\bf r})\times
\hat{{\bf B}}({\bf r'})
\rangle _{{\bf r'}\to {\bf r}}
\end{equation}
on the body's internal charge and current densities
\begin{gather}
\label{rhoint}
\underline{\hat{\rho}}({\bf r},\omega)=\frac{i\omega}{c^2}\bm{\nabla}
\cdot
\int \D ^3r'\, \ten{G}({\bf r}, {\bf r'}, \omega)
\cdot \underline{\hat{{\bf j}}}_N({\bf r'},\omega),\\
\label{jint}
\underline{\hat{{\bf j}}}({\bf r},\omega)
 =\Bigl(\bm{\nabla}\times\bm{\nabla}\times-
 \frac{\omega^2}{c^2}\Bigr)
 \int \D ^3r'\,\ten{G}({\bf r}, {\bf r'},\omega)\cdot
\underline{\hat{{\bf j}}}_N({\bf r'},\omega).
\end{gather}
Note that the coincidence limit ${\bf r'}\to {\bf r}$ has to be
performed such that (divergent) self-forces are discarded.

On using Eq.~(\ref{jn}), the bosonic commutation relations for
$\hat{\mathbf{f}}$, $\hat{\mathbf{f}}^\dagger$, and the definition of
the vacuum state, we find the following nonvanishing expectation
values:
\begin{align}
\label{jn1}
\langle\underline{\hat{{\bf j}}}_N({\bf r},\omega)
 \underline{\hat{{\bf j}}}^\dagger_N({\bf r'},\omega')\rangle
=&\hspace{0.1cm} \hbar\omega^2\varepsilon_0\pi^{-1}
\delta(\omega-\omega')
\varepsilon_I(\vec{r},\omega)\nonumber\\
& \times  \bm{\delta}(\vec{r}-\vec{r'})
 \Theta[\varepsilon_I(\vec{r},\omega)],
\\
\label{jn2}
\langle\underline{\hat{{\bf j}}}_N^\dagger({\bf r},\omega)
 \underline{\hat{{\bf j}}}_N({\bf r'},\omega')\rangle
=&-\hbar\omega^2\varepsilon_0\pi^{-1}\delta(\omega-\omega')
\varepsilon_I(\vec{r},\omega)\nonumber\\
&\times  \bm{\delta}(\vec{r}-\vec{r'})
\Theta[-\varepsilon_I(\vec{r},\omega)].
\end{align}
We evaluate the force by combining Eqs.~(\ref{edef}),(\ref{bdef}),
(\ref{e20})--(\ref{jn2}) and writing
${\bf a}\times{\bf b}%
=-{\rm Tr}(\ten{I}\times{\bf a}\otimes{\bf b})$
($[{\rm Tr}\ten{T}]_i=T_{kik}$) for the
$\hat{{\bf j}}
\times\hat{{\bf B}}$ term. Eliminating
$\Theta[\varepsilon_I(\vec{s},\omega)]$ according to
$\Theta[\varepsilon_I(\vec{s},\omega)]=1
-\Theta[-\varepsilon_I(\vec{s},\omega)]$,
those spatial integrals not depending on
$\Theta[-\varepsilon_I(\vec{s},\omega)]$ can be performed via
\begin{equation}
\frac{\omega^2}{c^2}\int \D^3 s\,
\varepsilon_I({\bf s},\omega)\ten{G}({\bf r},{\bf s},\omega)
\cdot
\ten{G}^\ast({\bf s},{\bf r'},\omega)
=\Im \ten{G}({\bf r},{\bf r'},\omega),
\end{equation}
so the Casimir force can be given by
${\bf F}={\bf F}^\mathrm{r}+{\bf F}^\mathrm{nr}$
with
\begin{align}
\label{F1}
&{\bf F}^\mathrm{nr}
=\frac{\hbar}{\pi}\int_{0}^\infty\dif\omega\int_{V}\dif^3r\,
 \Bigl\{\frac{\omega^2}{c^2}\,\bm{\nabla}\sprod
 \mathrm{Im}\ten{G}(\vect{r},\vect{r'},\omega)\nonumber\\
&+\trace\Bigl[\ten{I}\!\vprod\!
 \biggl(\bm{\nabla}\!\vprod\!\bm{\nabla}\!\vprod\,
 -\frac{\omega^2}{c^2}\Bigr)
 \mathrm{Im}\ten{G}(\vect{r},\vect{r'},\omega)\!\vprod\!
 \overleftarrow{\bm{\nabla}}'\Bigr]
 \biggr\}_{\vect{r'}\to\vect{r}}
\end{align}
and
\begin{align}
\label{F2}
{\bf F}^\mathrm{r}
=&-\frac{2\hbar}{\pi c^2}\!
\int _{V}\! \D^3r\!\int_0^\infty\!\!\D \omega \omega^2\!
  \int \D^3 s\varepsilon_I(\vec{s},
\omega) \Theta[-\varepsilon_I(\vec{s},\omega)]
 \nonumber\\
&\times\Re\bigl\{\omega^2/c^2\bm {\nabla}
\cdot\ten{G}({\bf r},{\bf s},\omega)
  \cdot
\ten{G}^\ast({\bf s},{\bf r'},\omega)\nonumber\\
&\hspace{7ex}+{\rm Tr}\bigl[\ten{I}\times
\bigl(\bm{\nabla}\!\times\!\bm{\nabla}\!\times\!
-\omega^2/c^2\bigr)
\ten{G}({\bf r},{\bf s},\omega)\nonumber\\
&\hspace{13ex}\cdot\ten{G}^\ast({\bf s},{\bf r'},\omega)
\times\overleftarrow{\bm{\nabla}}'\bigr]
\bigr\}_{{\bf r'}\rightarrow {\bf r}}.
\end{align}

Equations~(\ref{F1}) and (\ref{F2}) represent general expressions
for the Casimir force acting on a linearly polarisable body of
arbitrary shape and material in an arbitrary environment of additional
bodies or media, where any of the bodies may be amplifying. The term
${\bf F}^\mathrm{nr}$ is a purely nonresonant Lifshitz-type
contribution to the force. It can be rewritten as an integral over
purely imaginary frequencies $\omega=i\xi$ and has exactly the same
form as for purely absorbing bodies. In Ref.~\cite{Leonhardt2007}, the
nonresonant term ${\bf F}^\mathrm{nr}$ is identified with the total
Casimir force and it  is shown that ${\bf F}^\mathrm{nr}$ may become
repulsive in the presence of amplifying media as a consequence of the
property $\varepsilon(i\xi)\le 1$. The resonant term 
${\bf F}^\mathrm{r}$ has never been given before. It only arises in
the presence of amplifying bodies, in which case it can dominate the
total Casimir force. As evident from the factor
$\Theta[-\varepsilon_I(\vec{s},\omega)]$, the force component
${\bf F}^\mathrm{r}$ is associated with emission processes [the
emission spectrum being related to $-\varepsilon_I(\vec{s},\omega)$].
%
%%%%%%%%%%%%%%%%%%%%%%%%%%%%%%%%%%%%%%%%%%%%%%%%%%%%%%%%%%%%%%%%%%%%%%
%
\section{Contact to Casimir--Polder forces}
We will next establish a relation between the Casimir force 
${\bf F}={\bf F}^\mathrm{r}+{\bf F}^\mathrm{nr}$
[with ${\bf F}^\mathrm{r}$ and ${\bf F}^\mathrm{nr}$ being given by
Eqs.~(\ref{F1}) and (\ref{F2}), respectively] and the well-understood
CP force on excited atoms. In this way, we will be able to
substantiate the role of emission mentioned above. To that end, we
consider the Casimir force on an optically dilute amplifying body of
volume $V$ placed in a free-space region in an environment of purely
absorbing bodies. We follow the procedure outlined in
Ref.~\cite{Raabe2006} for an absorbing dielectric body.

We begin with the nonresonant force component
${\bf F}^\mathrm{nr}$, Eq.~(\ref{F1}), and explicitly introduce
the electric susceptibility
$\chi({\bf r},\omega)=\varepsilon({\bf r},\omega)-1$ (${\bf r}\in V$)
of the body by invoking the relations
\begin{gather}
\label{eq31}
\Bigl(\bm{\nabla}\!\times\!\bm{\nabla}\!\times
-\frac{\omega^2}{c^2}\Bigr)
 \Im\ten{G}({\bf r},{\bf r}',\omega)
\!=\!\frac{\omega^2}{c^2}\,
 {\rm Im}[\chi({\bf r},\omega)
 \ten{G}({\bf r},{\bf r'},\omega)],\\
\bm{\nabla}\cdot\Im\ten{G}({\bf r},{\bf r}',\omega)
=-{\rm Im}[\bm{\nabla}\cdot
\chi({\bf r},\omega)\ten{G}({\bf r},{\bf r'},\omega)],
\end{gather}
which follow from Eq.~(\ref{e4}). Expanding the resulting expression
for ${\bf F}^\mathrm{nr}$ via ${\rm Tr}\bigl[\ten{I}\times\ten{G}%
\times\overleftarrow{\bm{\nabla}}'\bigr]%
=\bm{\nabla}'{\rm Tr}\ten{G}-\bm{\nabla}'\cdot\ten{G}$
and exploiting the fact that terms involving a total divergence can
be converted to a surface integral that vanishes for a body in free
space, one obtains
\begin{multline}
\label{F1dilute}
{\bf F}^\mathrm{nr}
=\frac{\hbar}{2\pi}\int_{V}\dif^3r\int_{0}^\infty\dif\omega\,
 \frac{\omega^2}{c^2}\\
\times
 \Im[\chi({\bf r},\omega)\bm{\nabla}
 {\rm Tr}\ten{G}^{(1)}(\vect{r},\vect{r},\omega)],
\end{multline}
where the symmetry $\ten{G}(\vect{r}',\vect{r},\omega)%
=\ten{G}^\mathrm{T}(\vect{r},\vect{r}',\omega)$ of the Green
tensor has been used. In addition, we have assumed the body to be
homogeneous and performed the coincidence limit by simply replacing
the Green tensor with its scattering part $\ten{G}^{(1)}$ (see the
discussion in Ref.~\cite{Raabe2006}). Next, we exploit the fact that
the amplifying body is optically dilute and expand 
${\bf F}^\mathrm{nr}$ as given by Eq.~(\ref{F1dilute}) to leading
(linear) order in the susceptibility $\chi({\bf r},\omega)$ of the
amplifying body. We thus have to replace $\ten{G}$ with its zero-order
approximation $\overline{\ten{G}}$, i.e., the Green tensor of the
system in the absence of the amplifying body which is the solution to
the Helmholtz equation~(\ref{e4}) with
\begin{equation}
\label{epsilonmubar}
\overline{\varepsilon}({\bf r},\omega)
=\begin{cases}
\varepsilon({\bf r},\omega)\quad\mbox{for }
 {\bf r}\notin V,\\
1 \quad\mbox{for }{\bf r}\in V.
\end{cases}
\end{equation}
in place of $\varepsilon({\bf r},\omega)$. Finally, we assume that the
amplifying body consists of a gas of isotropic atoms in an excited
state $|n\rangle$ with polarizability
\begin{equation}
\label{alpha}
\alpha_n(\omega)=\lim _{\epsilon\rightarrow 0} \frac{1}{3\hbar}
\sum_k\biggl[
\frac{|\mathbf{d}_{nk}|^2}{\omega+\omega_{kn}+i\epsilon}
-\frac{|\mathbf{d}_{nk}|^2}{\omega-\omega_{kn}+i\epsilon}\biggr]
\end{equation}
($\omega_{kn}$: transition frequencies, $\mathbf{d}_{nk}$: electric
dipole matrix elements), which can be related to the electric
susceptibility of the body via the linearised Clausius--Mossotti law
$\chi(\omega)=\varepsilon_0^{-1}\eta\alpha_n(\omega)$ ($\eta$: atomic
number density). Transforming the frequency integral to the positive
imaginary axis, we obtain
${\bf F}^{\rm nr}=-\int\D^3r\,\eta\bm{\nabla}U_n^{\rm nr}({\bf r})$,
where 
\begin{equation}
\label{F1od}
U_n^{\rm nr}({\bf r})
=\frac{\hbar\mu_0}{2\pi}\int _0^\infty \D\xi\,
\xi^2\alpha_n(i\xi)
{\rm Tr}\overline{\ten{G}}{}^{(1)}(\vec{r},\vec{r},i\xi)
\end{equation}
is the nonresonant CP potential of the excited body atoms
\cite{Wylie1985}. It should be pointed out that there is an important
difference to the case of the force on an absorbing object made of
ground-state atoms: In the latter case, all of the frequencies
$\omega_{kn}$ in Eqs.~(\ref{alpha}) are positive so that the
respective (virtual) transitions contribute to the nonresonant CP
potential with the same sign. For excited atoms, upward as well as
downward transitions are possible, so that positive and negative
$\omega_{kn}$ occur and the overall sign of the nonresonant force can
be reversed to make it repulsive.

Let us next consider the resonant force component 
${\bf F}^\mathrm{r}$, Eq.~(\ref{F2}), following similar steps as
above. The linear approximation of ${\bf F}^\mathrm{r}$ can be
obtained by using the zero-order approximation to Eq.~(\ref{e4})
together with the identity
$\omega^2/c^2\Nabla\cdot\ten{G}({\bf r},{\bf r}',\omega)%
=-\Nabla\delta({\bf r}-{\bf r}')$ and replacing $\ten{G}^\ast$ with
$\overline{\ten{G}}{}^\ast$. Expanding the result 
according to ${\rm Tr}\bigl[\ten{I}\times\ten{G}%
\times\overleftarrow{\bm{\nabla}}'\bigr]%
=\bm{\nabla}'{\rm Tr}\ten{G}-\bm{\nabla}'\cdot\ten{G}$
and discarding, for a body in free space, terms involving total
divergences we derive
\begin{multline}
\label{Fod4}
{\bf F}^\mathrm{r}
=-\frac{\hbar}{\pi}\int_{V}\dif^3r
\int_{0}^\infty\dif\omega\,\frac{\omega^2}{c^2}\,
 \Theta[-\varepsilon_I({\bf r},\omega)]
 \varepsilon_I({\bf r},\omega)\\
\times\bm{\nabla}
 {\rm Tr}\,\Re
 \overline{\ten{G}}^{(1)}(\vect{r},\vect{r},\omega),
\end{multline}
where we have again assumed the body to be homogeneous and performed
the coincidence limit by replacing the Green tensor with its
scattering part. Relating $\varepsilon_I$ to the polarizability of the
atoms by means of the linearised Clausius--Mossotti relation, we
finally obtain ${\bf F}^\mathrm{r}=-\int \D ^3 r\,\eta\bm{\nabla}
U_n^{\rm r}({\bf r})$, where
\begin{multline}
\label{eq48}
U_n^\mathrm{r}({\bf r})=
\frac{\hbar\mu_0}{\pi}\int_{0}^\infty\dif\omega\,\omega^2
 \Theta[-\Im\alpha_n(\omega)]
 \Im\alpha_n(\omega)\\
\times
 {\rm Tr}\,\Re
 \overline{\ten{G}}^{(1)}(\vect{r},\vect{r},\omega)
\end{multline}
is nothing but the resonant part of the CP potential of the excited
atoms contained in the body. By using the relation
\begin{equation}
\label{Imalpha}
 \alpha_I(\omega)=\frac{\pi}{3\hbar}\sum_k |{\bf d}_{nk}|^2
       [\delta(\omega+\omega_{nk})-\delta(\omega-\omega_{nk})],
\end{equation}
which follow from definition~(\ref{alpha}) together with the identity
$\lim_{\epsilon\to 0}1/(x+i\epsilon)=\mathcal{P}/x-i\pi\delta(x)$
($\mathcal{P}$: principal value), we can write it in the more familiar
form \cite{Wylie1985}
\begin{equation}
\label{Fod6}
U_n^\mathrm{r}({\bf r})
=-\frac{\mu_0}{3}\sum_{k}\Theta(\omega_{nk})
\omega_{nk}^2|{\bf d}_{nk}|^2
{\rm Tr Re} \overline{\ten{G}}^{(1)}({\bf r},{\bf r}, \omega_{nk}).
\end{equation}

Combining the results for $\vec{F}^\mathrm{nr}$ and
$\vec{F}^\mathrm{r}$, we see that the Casimir force on an optically
dilute, homogeneous, amplifying electric body is the sum of the CP
forces on the excited atoms contained therein,
\begin{equation}
\label{eq52}
 {\bf F}
=\vec{F}^\mathrm{nr}+\vec{F}^\mathrm{r}
=-\int \D ^3 r
\eta
\Nabla U_n({\bf r})
\end{equation}
[$U_n({\bf r})=U_n^\mathrm{nr}({\bf r})+U_n^\mathrm{r}({\bf r})$].
This result generalises similar findings for purely absorbing bodies
(consisting of ground-state atoms) \cite{Tomas2005b, Raabe2006,
Buhmann2006a, Buhmann2006b} to amplifying ones. In particular, the
nonresonant and resonant components of the Casimir force,
Eqs.~(\ref{F1}) and (\ref{F2}), are directly related to the respective
CP-potential terms which in turn are associated with virtual and real
transitions of the atoms. Recall that for an atom in front of a plate
at zero temperature, the nonresonant CP potential is proportional to
$1/z^3$ and $1/z^4$ in the nonretarded and retarded limits ($z$:
atom--plate separation), while the resonant potential as governed by a
$1/z^3$ power law in the nonretarded regime makes way for a spatially
oscillating $\cos(2\omega_{nk}z/c)/z$ asymptote for retarded distances
\cite{Barton70}. As we have seen, the most important difference
between forces on amplifying as opposed to absorbing bodies is the
presence of a strong, resonant force contribution which is associated
with real transitions of the excited body atoms and hence with
emission processes of the body. 
%
%%%%%%%%%%%%%%%%%%%%%%%%%%%%%%%%%%%%%%%%%%%%%%%%%%%%%%%%%%%%%%%%%%%%%%
%
\section{Summary and perspective}
On the basis of the consistent framework provided by macroscopic QED,
we have developed an exact theory of Casimir forces in arrangements
of linearly responding, electrically polarisable bodies of arbitrary
shape, with special emphasis on amplifying bodies. The formulas
(\ref{F1}) and (\ref{F2}) show that the Casimir force can be
decomposed into two parts: A nonresonant Lifshitz-type component that
looks formally the same as in the case of a purely absorbing body and
a novel, resonant component that is a direct consequence of the
amplification in the system and contains an integration over bodies
and frequencies where the imaginary part of the electric permittivity
is negative. We have demonstrated that in the dilute-gas limit,
the Casimir force on an amplifying body in the presence of absorbing
bodies is given by a sum of CP forces over the excited body atoms;
our theory is hence the natural generalisation of additive approaches
beyond their scope.

Two points to be addressed in more detail in the future are (i) the
finite-temperature case and (ii) the relation of the excited
atom--body force to atom--atom forces. (i)~Finite temperature of the
absorbing environment may be accounted for by introducing a thermal
density matrix $\hat{\rho}_T=%
\exp[-\hat{H}_+/(k_\mathrm{B}T)]/%
\trace\{\exp[-\hat{H}_+/(k_\mathrm{B}T)]\}$. This leads to a factor
$[2n_T(\omega)\!+\!1]$ with thermal photon number $n_T(\omega)$ in
Eq.~(\ref{F1}), so the $\xi$-integral in Eq.~(\ref{F1od}) will be
replaced with a Matsubara sum, cf.~Ref.~\cite{Buhmann2008t}. The 
$T\!\to\!0$
limit of the nonresonant force being a nonuniform asymptotic expansion
\cite{Ninham98,Wennerstroem99}, the $T\!=\!0$ results are only valid
for distances $z\!\ll\!\hbar c/(2\pi k_\mathrm{B}T)$. (ii)~The
theoretically and experimentally established resonant and spatially
oscillating forces between excited atoms and bodies are in contrast
with the non-oscillating atom--atom forces found in the majority of
theoretical works \cite{Bostroem2003,Sherkunov06}. The discrepancy
indicates that either the inclusion of atomic linewidth at the heart
of excited atom--atom calculations has to be carefully reconsidered 
\cite{Buhmann04} or that microscopic atom--atom interactions are
fundamentally different from the collective atom--body interaction.

Our results, which can be extended to magnetoelectric, anisotropic
or nonlocally responding media in a straightforward way, present a
reliable framework for enhancing the impact of metamaterial properties
such as negative refraction on dispersion forces or realizing
repulsive forces. In particular, perfect lens scenarios can be
investigated by using the appropriate Green tensors. Note that the
strategy employed in this Rapid Communication can also be employed to
study CP forces on atoms in the presence of amplifying bodies, where a
rich dynamics associated with the exchange of excitations is to be
expected.
%
%%%%%%%%%%%%%%%%%%%%%%%%%%%%%%%%%%%%%%%%%%%%%%%%%%%%%%%%%%%%%%%%%%%%%%
%
\acknowledgments
The work was supported by Deutsche Forschungsgemeinschaft. We
gratefully acknowledge funding from the Alexander von Humboldt
Foundation (S.Y.B.) and the Vietnam Education Foundation (H.T.D.).
We acknowledge fruitful discussions with C. Raabe and M. Fleischhauer. 

\end{document}